\RequirePackage{fix-cm}
\documentclass[smallextended]{svjour3}       
\smartqed  
\usepackage{graphicx}
\usepackage{amsmath,amssymb,amscd,latexsym}
\usepackage{paralist}
\usepackage{verbatim}
\usepackage{epsfig}
\usepackage[colorlinks=true]{hyperref}
\usepackage{color,soul,verbatim}
\usepackage{epstopdf}
\usepackage{longtable}
\usepackage{array,arydshln,tabu}
\usepackage{lipsum}
\usepackage{ntheorem}
\newtheorem*{theorem-non}{Theorem}
\newtheorem*{conjecture-non}{Conjecture}
\newcommand{\Fq}{\mathbb{F}_q}
\usepackage{soul,color,xcolor}

\usepackage{algorithm}
\usepackage[noend]{algpseudocode}
\begin{document}

\title{A New Algorithm for Equivalence of Cyclic Codes and Its  Applications
}


\author{Nuh Aydin         \and
        R. Oliver VandenBerg 
}


\institute{N. Aydin \at
              Department of Mathematics and Statistics, Kenyon College, Gambier OH, USA 43022 \\
              Tel.: +1-740-4275674\\
              Fax: +1-740-4275573\\
              \email{aydinn@kenyon.edu}           
           \and
           R. Oliver VandenBerg \at
              Department of Mathematics and Statistics, Kenyon College, Gambier OH, USA 43022\\
							\email{vandenberg1@kenyon.edu}
}

\date{Received: date / Accepted: date}

\maketitle

\begin{abstract}
Cyclic codes are among the most important families of codes in coding theory for both theoretical and practical reasons. Despite their prominence and intensive research on cyclic codes for over a half century, there are still open problems related to cyclic codes.  In this work, we use recent results on the equivalence of cyclic codes to create a more efficient algorithm to partition cyclic codes by equivalence based on cyclotomic cosets. This algorithm is then implemented to carry out computer searches for both cyclic codes and quasi-cyclic (QC) codes with good parameters. We also generalize these results to repeated-root cases. We have found several new linear codes that are cyclic or QC as an application of the new approach, as well as more desirable constructions for linear codes with best known parameters. With the additional new codes obtained through standard constructions, we have found a total of 14 new linear codes.
\end{abstract}

\section{Introduction}
\label{intro}
We study the problem of determining equivalence between cyclic codes in this work. There are several reasons that motivate this study. The class of cyclic codes is one of the oldest families in coding theory that have been studied for over sixty years \cite{Oldcyc1,Oldcyc2}. They are important for both theoretical and practical purposes and they establish a fundamental link between coding theory and algebra. There has been intensive research on codes over rings in the past few decades. Whenever a new type of ring is introduced to coding theory, one of the very first things researchers do is to study cyclic codes over that ring.  In addition to the fact that both cyclic codes and code equivalence are fundamental topics in coding theory, there are additional benefits that may be obtained from new results on this question as explained below. 

A linear code $C$ over a finite field $\Fq$ is a vector subspace of $\Fq^n$. Each code has a length ($n$), dimension ($k$), and minimum Hamming distance ($d$). Constructing codes with optimal parameters is one of the most important and challenging problems in coding theory. This optimization problem can be formulated in a few different ways. For example, given the length $n$ and the dimension $k$ over a certain finite field $\Fq$, we look for the largest possible value $d_q(n,k)$ of the minimum distance. Theoretical upper bounds are available for $d_q(n,k)$, and codes whose parameters attain an upper bound are called optimal codes, sometimes $d$-optimal codes. The online database \cite{tables} gives information about best known linear codes (BKLC) over small finite fields  $\Fq$, $q\le 9$, including lower and upper bounds on $d_q(n,k)$. Lower bounds are usually obtained by explicit constructions. We observe from the database that in most cases there are gaps between lower and upper bounds on $d_q(n,k)$. In general, optimal codes are known only when either $k$ or $n-k$ is small.  

Computer searches are often used  to find linear codes with best known parameters, however there are inherent limitations to the computational approach. First, determining the minimum distance of a linear code is computationally intractable \cite{NPhard} so it takes significant time to find the minimum distance of a single code when the dimension is large. Second, for a given length and dimension, the number  $\displaystyle{\frac{(q^n-1)(q^n-q)\cdots (q^n-q^{k-1})}{(q^k-1)(q^k-q)\cdots (q^k-q^{k-1})}}$ of linear codes is very large  and grows quickly. Hence, exhaustive computer searches on linear codes is not feasible. Therefore, we  focus on specific classes of codes with rich mathematical structures that are known to contain many codes with good parameters. One such family of codes is the class of quasi-twisted (QT) codes that contains cyclic, constacyclic and quasi-cyclic (QC) codes as special cases. Hundreds of record breaking codes have been obtained in the last few decades by computer searches within these classes (see \cite{NewTernary},\cite{NewConsta},\cite{BinaryQC},\cite{NewGF7}, \cite{QT5} for a few examples) all of which are generalizations of cyclic codes. Our work in this paper shows that new results on equivalence of cyclic codes can be useful for computational purposes.

Another area of research that has been receiving much attention recently is  code-based cryptography due to its promise for the age of quantum computers. One specific problem of great interest in the field is to reduce the key size in the McEliece cryptosystem \cite{MECS}, one of the earliest examples of public key cryptosystem --as old as the widely used RSA cryptosystem-- that did not become widely used due to its large key size. However, given the promise of the McEliece system for post-quantum cryptography there has been a renewed interest on the subject. Cyclic codes and their various generalizations, such as QC codes, are being considered in an effort to reduce the key size in the  McEliece cryptosystem (see \cite{QCMECS} for a recent example). Code equivalence is of fundamental importance in code-based cryptography which provides another motivation for this work.

\section{Basic Definitions}
\label{sec:1}

A linear code $C$ is called cyclic if whenever $c = (c_0,c_1,...,c_{n-1})$ is a codeword, then so is its cyclic shift $\pi(c)=(c_{n-1},c_0,c_1,...,c_{n-2})$. If we represent a vector  $c = (c_0,c_1,...,c_{n-1})$ by the polynomial $c(x) = c_0+c_1x+\cdots +c_{n-1}x^{n-1}$, then we obtain a vector space isomorphism between $\Fq^n$ and the set of all polynomials over $\Fq$ of degree less than $n$. In this representation, the cyclic shift of $c$ in $\Fq^n$  corresponds to the multiplication of $c(x)$ by $x$ in the quotient ring $\Fq[x]/\langle x^n-1 \rangle$. It is well known that  cyclic codes of length $n$ over $\Fq$  are precisely the ideals of the  ring $\Fq[x]/\langle x^n-1 \rangle$ which is a principal ideal ring (a ring in which every ideal is generated by a single element) and there is a one-to-one correspondence between cyclic codes of length $n$ over $\Fq$ and divisors of $x^n-1$ over $\Fq$. In general, a cyclic code $C$  has many generators. Among all generators of $C$, there is a  generator polynomial $g(x)$ that is uniquely determined by the following two conditions (i) $g(x)$ is monic (ii) $g(x)$ is a non-zero polynomial of smallest degree in $C$. We write $C=\langle g(x) \rangle$. This unique generator must divide $x^n-1$, and it is called the (standard) generator of $C$. Hence, we can write $x^n-1=g(x)h(x)$ and call $h(x)$ the check polynomial of $C$. Moreover, any other generator of $C$ is of the form $g(x)f(x)$ where $\gcd(f(x),h(x))=1$ \cite{QTOriginal}. A cyclic code may be defined either by its generator polynomial or its check polynomial.

Given the generator polynomial $g(x)=g_0+g_1x+\cdots+g_tx^t$ of a cyclic code $C$, we obtain a generator matrix for $C$ as a circulant matrix of the form 

\begin{center}
\[
\left[
\begin{array}{ccccccc}
g_0&g_1&\cdots&g_{t}& 0 & \cdots &0\\
0&g_0&g_1&\cdots&g_{t}& 0 \cdots&0\\
\hdots \\
0& \hdots &0 &g_0& g_1& \cdots &g_{t} \\

\end{array}
\right].
\]
\end{center}

\noindent where each row is the cyclic shift of the row above it. The dimension of $C$ is $n-\deg(g(x))$.

A fundamental notion in the study of cyclic codes is that of cyclotomic coset, a subset of 
$\mathbb{Z}_n$.
\begin{definition}
Let $\gcd(n,q)=1$. For any $s \in \mathbb{Z}_n$,  the $q$-cyclotomic coset of $n$ containing $s$ is the set $S_s=\{sq^j \mod n :  j= 0,1,2...\}$
\end{definition}

Assuming $\gcd(n,q)=1$, there is a one-to-one correspondence between the irreducible factors of $x^n-1$ and cyclotomic cosets mod $n$. Therefore there is a correspondence between cyclic codes and unions of cyclotomic cosets when $\gcd(n,q)=1$. This correspondence is obtained via the  $n^{th}$ roots of unity. Let $\alpha \in \mathbb{F}_{q^t}$ be a primitive $n^{\text th}$ root of unity over $\Fq$. Then,  $\displaystyle{x^n-1=\prod_{i=0}^{n-1} (x-\alpha^i)}$ in $\mathbb{F}_{q^t}[x]$. Each cyclotomic coset $S_s=\{s,sq,sq^2,...,sq^{r-1}\} \mod n$ corresponds to an irreducible divisor $p(x)$ of $x^n-1$ over $\Fq$ by $p(x)=\displaystyle{\prod_{i \in S_s}(x-\alpha^i)}$. We denote the cyclotomic coset corresponding with a given polynomial $g$ by $S_g$.

Equivalence of codes is a central concept for this work. Two linear codes are called  equivalent if  one can be obtained from the other by any combination of the following transformations 
\begin{enumerate}
    \item  A permutation of the coordinates. 
    \item Multiplication of elements in a fixed position by a non-zero scalar in $\Fq$. 
    \item Applying a field automorphism  of $\Fq$ to each component of the vectors. 
\end{enumerate}
 If only (1) is used, then the codes are called permutation equivalent. This is a very important special case. We can summarize all of these conditions in the following way.

\begin{definition} 
Two linear codes $C_1,C_2\subseteq \mathbb{F}_q^n$ are equivalent if there exists a monomial matrix $M$ and an automorphism $\phi$ over $\Fq$ such that $C_1=C_2M\phi$. 
\end{definition}

It is well known that equivalent codes have the same parameters. This has important implications for computer searches because a search algorithm need not examine any code that is  equivalent to codes which have been already searched. This work generalizes our previous results in \cite{genASR} on the equivalence of cyclic codes and their generalization to constacyclic codes. Further, it includes a new algorithm for partitioning cyclic codes by equivalence and a search algorithm that only checks the minimum distance of codes that are not equivalent to any of the codes already examined.

\section{Extending Results on Equivalence of Cyclic Codes to Repeated Root Case}
\label{sec:2}
Under the assumption of $\gcd(n,q)=1$, the following theorems are proven in \cite{genASR}. 
\begin{theorem}\cite{genASR}
 \label{theorem:CycThm1}
Let $g_1(x)$ and $g_2(x)$ be the standard generators of cyclic codes of length $n$ over $\Fq$ and assume $\gcd(e,n)=1$. Then the isometry $$\Phi:\Fq[x]/\langle x^n-1 \rangle \longrightarrow \Fq[x]/\langle x^n-1 \rangle$$ given by $$x \mapsto x^{e} \mod x^{n}-1$$ has the property $g_2(x)=\phi(g_1(x)))$  if and only if the map   $\phi: S_{g_1} \mapsto S_{g_2}$ given by $\phi(z)=e^{-1}z   \mod  n$, where $e^{-1}$ is the multiplicative inverse of $e \mod n$,  is a bijection.
\end{theorem}
\begin{theorem}\cite{genASR}
\label{theorem:CycThm2}
Let $g_1(x)$  be the standard generator of a cyclic code of length $n$ over $\Fq$ where $\gcd(n,q)=1$, and let  $\delta=\alpha^{-b}$ where $\alpha$ is a primitive $n^{\text th}$ root of unity, such that $n$ divides $b\cdot \deg(g_1(x))\cdot (q-1)$. Let $K$ be an extension field of $\Fq$ that contains $\delta$. Then the isometry $\displaystyle{\Phi:K[x]/\langle x^n-1 \rangle \mapsto K[x]/\langle x^n-1 \rangle}$ defined by $\displaystyle{\Phi(f(x))= f(\delta x) \mod(x^{n}-1)}$ has the property that $\Phi(g_1(x))\in \Fq[x]$ and  generates a cyclic code of length $n$ over $\Fq$ if and only if the map $\phi: \mathbb{Z}_n\mapsto \mathbb{Z}_n$ defined by $\phi(z)=z+b \mod n$ is a bijection such that $\phi(S_{g_1})=S_{\Phi(g_1)}$.
\end{theorem}

These theorems in their current form require that $x^n-1$ does not have repeated roots. This was assumed largely because there was not a construction for cyclotomic cosets in repeated-root cases. We now propose such a construction. First write the code length $n$ over a field with characteristic $q$ as $n=n_qq^i$ where $\gcd(q,n_q)=1$. Next we find the cyclotomic cosets $\mod n_q$.

Define a function $P$ which takes  cyclotomic cosets to polynomials. Let $\alpha$ be an $n_q^{th}$ root of unity, and $S$ be a cyclotomic coset $\mod n$. We define  $\displaystyle{P(S)=\prod_{i \in S}(x-\alpha^i)}$. Then we use a multiset to describe unions where if an irreducible factor of $x^n-1$ appears multiple times in a divisor, then the roots of that factor appear multiple times in the multiset. This means a multiset $MS$  is a union of not necessarily distinct cyclotomic cosets $S_1,S_2,...,S_k$ and it corresponds to the polynomial $P(MS)=P(S_1)\cdot P(S_2)\cdots P(S_k).$

For example, let us consider binary cyclic codes of length 14. Then $n_q=7$. So the cyclotomic cosets are $S_0=\{0\}$, $S_1=\{1,2,4\}$ and $S_3=\{3,5,6\}$. Now the following are  two possible multisets describing a union that corresponds to a divisor of $x^{14}-1: MS_1=\{1,2,4,1,2,4\}$ and $MS_2=\{3,5,6,3,5,6\}$. The corresponding codes to these multisets are equivalent  via a map between the sets of the form given in Theorem \ref{theorem:CycThm1} with $e^{-1}=5$. These correspond to the polynomials $P(MS_1)=P(S_1)P(S_1)=x^6+x^4+1$ and $P(MS_2)=P(S_3)P(S_3)=x^6+x^2+1$ which indeed generate equivalent codes.

Because the proofs of the Theorem \ref{theorem:CycThm1} and Theorem \ref{theorem:CycThm2} given in \cite{genASR} do not rely on the fact that the cyclotomic cosets have distinct elements, and use the same function to take cyclotomic cosets to polynomials,  both theorems also apply to this repeated-root construction as well. In generalizing the results in  Theorem \ref{theorem:CycThm1} and Theorem \ref{theorem:CycThm2} to the repeated root case, one needs to replace the condition $\gcd(e,n)=1$ by $\gcd(e,n_q)=1$. 

Moreover, this method can  be generalized to apply to constacyclic codes too.  This is because $x^n-a|x^{rn}-1$, where $r$ is the order of $a$ in the multiplicative group of $\Fq^{\ast}$.  The irreducible factors of $x^n-a$ over $\Fq$ correspond to a subset of cyclotomic cosets $\mod nr$. See \cite{QTOriginal} for more on this.

\section{A New Equivalence Testing Algorithm for Cyclic Codes}
The theorems in the previous section are the basis of an algorithm that we developed and implemented for testing cyclic code equivalence. This algorithm takes as input   the alphabet of the codes, the length of the codes, and two cyclotomic cosets (or unions of cyclotomic cosets) that define each cyclic code. The  output of the algorithm is a Boolean variable (called map in the pseudocode below) which indicates whether the two cyclic codes are equivalent. 
\vspace{-0.5 cm}
\begin{algorithm}[!h]
\caption{Equivalence Testing Algorithm for Cyclic Codes}\label{euclid}
\begin{algorithmic}[1]
\State Set alphabet, field, and code length ($q^m$,$F$, n) 
\State Enter cyclotomic cosets (gen1, gen2) of two cyclic codes.
\State map=false;
\State i=1;
\State \textbf{while} map eq 0 and i lt n \textbf{do}
\State \hspace{.7em}\textbf{if} GCD(i,n) eq 1 \textbf{then}
\State \hspace{.7em} \hspace{.7em} b=0;
\State \hspace{.7em}\hspace{.7em} \textbf{while} map eq 0 and b lt n \textbf{do}
\State \hspace{.7em} \hspace{.7em} \hspace{.7em} count=0;
\State \hspace{.7em} \hspace{.7em} \hspace{.7em}\textbf{for} j=1 to $\#$gen1 \textbf{do}
\State \hspace{.7em} \hspace{.7em} \hspace{.7em} \hspace{.7em}\textbf{if} Multiplicity(gen1,gen1[j]) eq Multiplicity(gen2,((gen2[j]*i+b) mod n)) \textbf{then}
\State \hspace{.7em} \hspace{.7em} \hspace{.7em} \hspace{.7em} \hspace{.7em} count=count+1;
\State \hspace{.7em} \hspace{.7em} \hspace{.7em} end \textbf{if}; end \textbf{for};
\State \hspace{.7em} \hspace{.7em} \hspace{.7em} \textbf{if} count eq $\#$gen1 \textbf{then}
\State \hspace{.7em} \hspace{.7em} \hspace{.7em} \hspace{.7em} map=true;
\State \hspace{.7em} \hspace{.7em} \hspace{.7em} end \textbf{if}; 
\State \hspace{.7em} \hspace{.7em}  b=b+1; end \textbf{while}; 
\State \hspace{.7em} end \textbf{if}; 
\State i=i+1; end \textbf{while};
\end{algorithmic}
\end{algorithm}

\vspace{-0.5 cm}

A few important remarks about this algorithm are in order. First, a cyclotomic coset does not uniquely determine an irreducible factor of $x^n-1$. It depends on the choice of the primitive $n^{th}$ root of unity. Nevertheless, different choices for the primitive $n^{th}$ root of unity will lead to equivalent codes. Secondly, since this algorithm is based on theorems that give sufficient conditions for two cyclic codes to be equivalent, when the algorithm returns that the two cyclic codes are equivalent, the output is definitely correct, and in this case the map between cyclotomic cosets is determined by the variables $i$ and $b$. However, when the output is false, it does not guarantee that the codes are not equivalent (except for the binary case. See section 7.) In that sense, it is similar to probabilistic primality testing algorithms that are true-biased. Despite this limitation, this algorithm is still very useful for our purposes as explained in the next section.    


\section{Computational Complexity of Cyclic Equivalence Test Algorithm}
\label{sec:3}

We consider the complexity of checking if there is a bijective map of the form $ex+b$ between the cyclotomic cosets corresponding with cyclic codes of length $n$. It is easy to see that the complexity is $O(n^3)$ since all possible combinations of $e,b<n$ can be tested (a total of $\leq n^2$  choices) to see if for each element of one cyclotomic coset is in the other cyclotomic coset at cost proportional to $n$.  This shows for cyclic codes testing equivalence using their cyclotomic cosets is a polynomial time algorithm in the length of the codes.

Determining whether two arbitrary linear codes are equivalent is much more challenging (\cite{Equiv4},\cite{Equiv7}, \cite{Crypt2}). It has been shown that testing code equivalence generally is equivalent to the graph isomorphism problem \cite{CheckEquiv}. For cyclic codes, using  cyclotomic cosets is a much more efficient test. It is particularly superior at large dimensions. This is because for a cyclic code of dimension $k$, the size of the cyclotomic coset is $n-k$, so it runs faster at higher dimensions, unlike other algorithms such as Magma \cite{magma} implementation (available using either IsEquivalent or IsIsomorphic command), which run slower. In comparing our algorithm with Magma's, it should be noted that (a) our algorithm is specifically for cyclic codes whereas Magma's algorithm is for general linear codes (b) given two cyclic codes, if our algorithm concludes that they are equivalent, then the result is certainly correct. However, when it does not make that conclusion, the codes may still be equivalent. 

Given these restrictions, it is not entirely fair to compare our algorithm with Magma's in general. However, for our purposes the comparison is justifiable because Magma does not offer any algorithm  specifically for cyclic codes (which is an easier case than general linear codes) and there are  many  cases where Magma's algorithm get stuck or takes so much time or memory that it is not useful. In fact, it was these shortcomings of Magma's algorithm that prompted us to come up with a more efficient algorithm for cyclic codes in the first place. Therefore, there are clear benefits and advantages of the new algorithm over existing algorithms of Magma, and a comparison is still relevant.  


The following table shows a comparison of Magma's IsEquivalent command and our algorithm on different generator polynomials in terms of CPU time and memory usage. In each case the generators were for codes of the same length and dimension, which were not necessarily equivalent. All of the test cases were executed on the online Magma calculator\footnote{http://magma.maths.usyd.edu.au/calc/} for consistency. 
\[
\begin{array}{p{.5cm} p{.5cm} p{.5cm} p{1cm} p{.75cm} p{1.5cm} p{1cm} p{1cm}}
&&&&CycCoset&&IsEquivalent\\
$q$&$n$&$k$&equiv&Time&Memory&Time&Memory\\
\rule{0pt}{3ex} 2&135&111&true&.04s&32MB&6.04s&32MB\\
\rule{0pt}{3ex} 3&80&61&true&.01s&32MB&$>$30s&*\\
\rule{0pt}{3ex} 3&80&19&true&.01s&32MB&.1s&32MB\\
\rule{0pt}{3ex} 5&72&14&false&.14s&32MB&.2s&32MB\\
\rule{0pt}{3ex} 5&72&18&false&.12s&32MB&$>$83s&*\\
\rule{0pt}{3ex} 7&18&12&true&.01s&32MB&.34s&32MB\\
\rule{0pt}{3ex} 7&36&28&false&.02s&32MB&3.490s&96.16MB\\
\rule{0pt}{3ex} 7&72&62&false&.03s&32MB&**&Unfinished\\
\rule{0pt}{3ex} 7&72&10&false&.14s&32MB&.2s&32MB\\

\end{array}
\]
* On these runs the Magma calculator's memory limit of 353MB was exceeded, and the time at which it was exceeded is given. \\
** On this run the 120s time limit was exceeded.

\section{Application to a  Search Algorithm}
\label{sec:4}

Our new  algorithm that partitions cyclic codes of a given length and dimension into equivalence classes is purely based on combinations of cyclotomic cosets. It generates all possible cyclic codes using the unions of cyclotomic cosets. Each new set is compared against all sets that have been found up to that point. If it is equivalent to a previous one then it will be discarded. Each  cyclotomic coset that is not found to be equivalent to any other is saved and the corresponding generator polynomial is obtained.  Computational evidence from testing  many different cases shows that this new algorithm  is both faster and less likely to get stuck than the Magma's algorithm. 

This partitioning algorithm has applications to both constacyclic searches and quasi-twisted (QT) searches. Earlier versions of the constacyclic search  algorithm (all versions before \cite{genASR}, such as \cite{NewConsta})  found the (standard) generators of all constacyclic codes of a given length one by one and computed the minimum distance for each code. Now, by first partitioning the constacyclic codes into equivalence classes, we only need to compute the minimum distance of one code from each equivalence class. Since computing the minimum distance of a linear code is computationally expensive \cite{NPhard}, the new approach  makes the search faster. Moreover, the new method of checking code equivalence based on cyclotomic cosets is much faster than  Magma's algorithm. This further speeds up the search over the previous algorithm \cite{genASR}. Taking advantage of these optimizations, our search produced 3 new cyclic codes with better parameters than previously best known liner codes \cite{tables}. It is  highly desirable but not very common (in fact, it is often impossible) that a best known code for a given parameter set has the cyclic structure.  Our search also produced many cyclic codes that have the same parameters as currently BKLCs in \cite{tables}. The constructions of currently BKLCs in \cite{tables} are frequently indirect, involving a long chain of steps. It would therefore be  desirable to obtain them as cyclic codes whenever possible.

One of our main motivations for this work was to speed up the search algorithm ASR. First introduced in \cite{QTOriginal}, ASR is a search algorithm for a particular type of 1-generator QT codes. It is based on Theorem 3.2 in \cite{QTOriginal}. It was then used in many subsequent works (e.g. \cite{QT3}, \cite{QT5}, \cite{BinaryQC}, \cite{NewGF7}, \cite{Ackerman2011}) and produced dozens of new linear codes.  We recently introduced a generalization of the ASR algorithm for searching QT codes \cite{genASR} which made made the original version of the ASR algorithm \cite{QTOriginal} more general by making use of the concept of code equivalence. In \cite{genASR}, we implemented the generalized algorithm using Magma software and found a number of new linear codes. We used Magma's IsEquivalent command to test equivalence of cyclic codes, which did not always work. For example, it experienced a bug in $GF(3)$. This bug was recently fixed, but still this command is often very slow and uses much more memory than the algorithm proposed here. Magma's algorithm is for checking equivalence between any two linear codes. We are not aware of another version that is specifically for cyclic codes.  The algorithm we propose here is specifically for cyclic codes and it uses our partitioning algorithm as a much faster way of obtaining the generators to be fed into the generalized ASR algorithm.  

We now present our partition algorithm which uses Algorithm 1 above. It takes as input the code length $n$, and the size $q^m$ of the finite field (code alphabet) which need not be relatively prime. As the output, it creates a list whose elements are coset representatives for each equivalence class of cyclic codes of length $n$ over $GF(q^m)$.  

\begin{algorithm}[!h]
\caption{Partition Algorithm for Cyclic Codes}\label{euclid}
\begin{algorithmic}[1]
\State Set alphabet, field, and code length ($q^m$,$F$, n) 
\State module=n; c1=0;
\State \textbf{while} (module mod q) eq 0 \textbf{do}
\State \hspace{.7em} module=module div q;
\State c1=c1+1; end \textbf{while};
\State cycreps=$q^{\hbox{c1}}$;
\State Generate all cyclotomic cosets (ocoset).
\State totalnum=(cycreps+1)$^{\#\hbox{ocosets}}-1$;
\State partit=[[]];
\State \textbf{for} i=1 to totalnum \textbf{do}
\State \hspace{.7em} Generate $i^{th}$ coset (poscos)
\State \hspace{.7em} Compare poscos to all cosets in partit using Algorithm 1.
\State \hspace{.7em} \textbf{if} it is not equivalent to any coset in partit \textbf{then}
\State \hspace{.7em} \hspace{.7em} Store poscos in partit
\State end \textbf{if}; end \textbf{for};
\end{algorithmic}
\end{algorithm}

Through an implementation of this algorithm we have been able to  search for QC codes over  $GF(3)$ which yielded 4 new record breaking codes. We were originally not able to carry out this search in \cite{genASR} due to a bug in Magma.  We point out that this algorithm can be modified for constacyclic codes and QT codes using the constacyclic cyclotomic coset construction given in \cite{QTOriginal}.

\section{A Limitation to the Partition}
We conjectured in \cite{genASR} (Conjecture 3) that all cyclic codes that are equivalent are equivalent through an affine map of the form $\phi(x)=ex+b$ on cyclotomic cosets.  We have found a few counterexamples to the conjecture for lengths which are multiples of 8 over $GF(3)$ and $GF(5)$. Consider the length 8 over $GF(3)$. The cyclotomic cosets are $\{0\},\{1,3\},\{2,6\},\{4\}$, and $\{5,7\}$. Two possible unions of cyclotomic cosets are $\{0,1,3,4\}$ and $\{1,2,3,6\}$. These turn out to correspond to equivalent codes even though no map of the form $ex+b$ exists between these cyclotomic cosets  (verified by exhaustive testing).

The following matrix, obtained by Magma command,  maps one code (its generator matrix) to the other.
\vspace{-0.7 cm}
\begin{center}
\[
\left[
\begin{array}{cccccccc}
1&0&0&0&0&0&0&0\\
0&1&0&0&0&0&0&0\\
0&0&0&0&0&0&2&0\\
0&0&0&0&0&0&0&2\\
0&0&0&0&1&0&0&0\\
0&0&0&0&0&1&0&0\\
0&0&2&0&0&0&0&0\\
0&0&0&2&0&0&0&0
\end{array}
\right].
\]
\end{center}

We observed that the structure of this matrix is similar to the counterexamples we have found for larger lengths and we are still looking for a pattern between the corresponding cyclotomic cosets.

The implication of this limitation for our search is that in some cases the partitions may have too many equivalence classes. These searches will still be exhaustive, but because of these counterexamples they may also have some redundancy in certain cases. Still, since the counterexamples seem to be rare, the method proposed is a more efficient partitioning algorithm. 

In the important special case of binary codes, the only notion of equivalence is permutation equivalence. More generally, it is shown in \cite{CyclicEquiv} that for cyclic codes of length $n$ over $\Fq$, the notions of permutation equivalence and  monomial equivalence coincide if and only if $\gcd(n,q-1)=1$. Over the binary field  all equivalent codes we have observed have a map of the form $x \rightarrow x^e$ between their generator polynomials. It turns out that this is a consequence of a result given in \cite{Berlekamp} (Theorem 5.81, page 142). Therefore, for binary codes the output of Algorithm 1 is certain (i.e., it is a full test) and it is a faster alternative to Magma's algorithm.  


\section{New Linear Codes}
\sectionmark{New Codes}\label{sec:results}

As a result of an implementation of this algorithm, we have found 3 new (``record breaking", i.e., codes with better parameters than currently BKLCs reported in \cite{tables}) linear codes that are cyclic. We like to point out that  it is rare for a record breaking code to be cyclic. Of the hundreds of new linear codes that have been discovered by computer searches over the last few decades, very few of them are cyclic. Since cyclic codes have a special structure that makes their implementation easier, it is desirable to have BKLCs as cyclic codes. This work presents one of the rare occasions where newly discovered linear codes are cyclic. Their parameters and generators are as follows: 

\begin{itemize}

\item $[146, 122, 9]_3$ cyclic code with generator \newline
$x^{24} + 2x^{23} + x^{22} + 2x^{21} + 2x^{20} + x^{19} + x^{16} + 2x^{14} + 
x^{13} + x^{11} + 2x^{10} + x^8 + x^5 + 2x^4 + 2x^3 + x^2 + 2x + 1$ \newline 
\item $[146, 121, 9]_3$ cyclic code  with generator \newline
$x^{25} + x^{24} + 2x^{23} + x^{22} + 2x^{20} + 2x^{19} + x^{17} + 2x^{16} + 2x^{15} + 2x^{14} + 2x^{13} + x^{12} + x^{11} + x^{10} + x^9 + 2x^8 + x^6 + x^5 + 2x^3 
+ x^2 + 2x + 2$ \newline
\item $[78, 63, 8]_5$ cyclic code  with generator \newline 
$x^{15} + 3x^{14} + x^{13} + x^{11} + 3x^9 + x^8 + 3x^7 + 4x^6 + 2x^5 + 2x^3 + x^2 + 4x + 4 $ 
\end{itemize}

We have also found 4  new ternary QC codes that are record breaking among linear codes. Their parameters and generators are listed below. They are all 1-generator QC codes of index 3 ($\ell=3$) with a generator of the form $(g(x),g(x)f_2(x),g(x)f_3(x))$. To represent a polynomial in a compact way, we just list the coefficients in ascending order of the terms. For example, the sequence $[2021]$ represents the polynomial $2+2x^2+x^3$.

\begin{longtabu}{>{$}l<{$} >{$}c<{$} >{$}c<{$} >{$}c<{$} @{} >{$}c<{$} @{} >{$}l<{$}}
\hline
[n,k,d]_q\qquad&\qquad&\multicolumn{3}{l}{Generator Polynomials}\\\nopagebreak
\hline
\endhead
\hline
\multicolumn{6}{c}{New QC Codes}\\\nopagebreak
\hline
\endfoot
{}[60,19,22]_3 & {}  g&=&[21]\\\nopagebreak
               &  f_2&=&[2200021200110200111]\\\nopagebreak
               &  f_3&=&[0012002212221102101]
\\[1ex]
{}[72,22,26]_3 & {}   g&=&[101]\\\nopagebreak
  &  f_2&{}={}&[1122220222021210022212]\\\nopagebreak
  &  f_3&{}={}&[1220021122022111]
\\[1ex]
{}[72,21,27]_3 &   g&=&[1221]\\\nopagebreak
  &  f_2&{}={}&[002100021111200121202]\\\nopagebreak
  &  f_3&{}={}&[200112121120102020202]
\\[1ex]
{}[72,18,29]_3 &   g&=&[1120221]\\\nopagebreak
  &  f_2&{}={}&[010110000212001001]\\\nopagebreak
  &  f_3&{}={}&[1210221200221001]
\\[1ex]
\end{longtabu}

Moreover, additional new codes are obtained from these codes using the standard constructions of shortening, extending, or puncturing a given linear code. The additional new codes we obtained have the following parameters:
[59,18,22], [70,19,27], [71,20,27], [71,21,26], [73,21,27], [73,22,27], [74,22,27]. With these, the total number of new linear codes we have obtained in this work is 14.

\subsection{Cyclic Codes  with the Same Parameters as BKLCs}

As noted above, in many cases the constructions of currently best known linear codes in \cite{tables} are  indirect that may involve multiple steps and manipulations. For example, the construction of the  binary  $[171,134,10]_2$ code in  \cite{tables} involves 4 steps starting with a BCH code of length 257, then  puncturing it, then extending the resulting code, then shortening the resulting code at many positions  and finally taking a subcode. The construction of the  $[71,55,8]_5$ code in  \cite{tables} requires 7 similar steps.  It would therefore be  desirable to obtain BKLCs as cyclic codes whenever possible. Our search  produced many cyclic codes that have the same parameters as currently BKLCs where the constructions of BKLCs in \cite{tables} are indirect with many steps. We present a subset of such codes here.

\begin{itemize}
\item $[171,134,10]_2$ cyclic code with generator \newline 
$[10010000001000110111101100010000001001]$ \newline

\item $[129,30,38]_2$ cyclic code with generator \newline 
$[1000100011101001001011100010001]$ \newline

\item $[126,100,8]_2$ cyclic code with generator\newline
$[111101111101111111011011011]$ \newline

\item $[56,42,6]_3$ cyclic code with generator \newline $[101201221201011]$ 
\item $[80,46,14]_3$ cyclic code with generator \newline $[12022102001111011100122212122201111]$\newline

\item $[80,60,8]_3$ cyclic code with generator\newline $[201120202000121220211]$\newline

\item $[60,53,4]_3$ cyclic code with generator \newline 
$[21001011]$

\item $[51,26,14]_4$cyclic code with generator \newline
$[1aa^2a^2aaa^211a^21a1101aaa0aa^211a^21]$ \newline

\item $[170,154,6]_4$ cyclic code with generator \newline
$[1a^21a11aa10a^211]$ \newline

\item $[71,55,8]_5$ cyclic code with generator \newline
$[12420401032243421]$ \newline

\item $[120,114,4]_5$ cyclic code with generator \newline
$[4424141]$ \newline

\item $[48,24,17]_7$ cyclic code with generator \newline
$[5025253510542120656461511]$ \newline

\item $[42,32,6]_7$ cyclic code with generator \newline
$[66541233451]$ 
\end{itemize}


%
%



\end{document}